\documentclass[twoside,12pt]{article}  %%% 这是 LaTeX 格式文件必须的指令，不可缺少
\usepackage{indentfirst}  %% 节标题后第一行段缩进
\usepackage{bm}    %%%%%  排印数学符号黑斜体时调用此宏包
\usepackage{graphicx}  %%%% 文中插入 eps 图形时调用此宏包，插图方法 1
\begin{document}    %% 文本文件开始，这是必须的指令

%\begin{CJK*}{GBK}{song}  %% 开始进入中文环境

%-------------------  First Head  -----------------------------------------
\thispagestyle{empty} \vspace*{0.8cm}\hbox
to\textwidth{\vbox{\hfill\huge\sf Chinese Physics B\hfill}}
\par\noindent\rule[3mm]{\textwidth}{0.2pt}\hspace*{-\textwidth}\noindent
\rule[2.5mm]{\textwidth}{0.2pt}

%=================== Text begin here =============================================

\begin{center}
%\LARGE\bf Cluster Monte Carlo simulations of the Eulerian bond-cubic model on the square lattice$^{*}$   %% 论文题目
\LARGE\bf Universal critical properties of the Eulerian bond-cubic model$^{*}$   %% 论文题目
\end{center}

\footnotetext{\hspace*{-.45cm}\footnotesize $^*$Project supported by National
Natural Science Foundation of China (Grant No 10675021), by the New Century
Excellent Talents in University (NCET). }
\footnotetext{\hspace*{-.45cm}\footnotesize $^\dag$Corresponding author. E-mail:  kjlis@bnu.edu.cn}
\footnotetext{\hspace*{-.45cm}\footnotesize $^\ddag$Corresponding author. E-mail:  yjdeng@ustc.edu.cn}

\begin{center}
\rm Ding Cheng-Xiang$^{\rm a)}$, \ \ Yao Gui-Yuan$^{\rm a)}$,\ \ Li Song $^{\rm b)\dagger}$,\\
  Deng You-Jin$^{\rm c)\ddagger}$, \ and  \ Guo Wen-An$^{\rm a)}$
\end{center}

\begin{center}
\begin{footnotesize} \sl
${}^{\rm a)}$ Physics Department, Beijing Normal University, Beijing 100875, China\\
${}^{\rm b)}$ Analysis and testing center, Beijing Normal University, Beijing 100875, China\\
${}^{\rm c)}$ Hefei National Laboratory for Physical Sciences at Microscale, Department of Modern Physics, University of Science and Technology of China, Hefei 230027, China
%%% 更多地址依次往下延续
\end{footnotesize}
\end{center}

\begin{center}
%\footnotesize (Received X XX XXXX; revised manuscript received X XX XXXX)
          %% (Received 日 月 年; revised manuscript received 日 月 年)
\end{center}

\vspace*{2mm}

\begin{center}
\begin{minipage}{15.5cm}
\parindent 20pt\footnotesize
We investigate the Eulerian bond-cubic model on the square lattice by means of Monte Carlo simulations, using an
efficient cluster algorithm and a finite-size scaling analysis.
The critical points and four critical exponents
%, including two fractal dimensions,
of the model are determined for several values of $n$.
Two of the exponents are fractal dimensions, which are obtained numerically for the first
time.
%by means of a finite-size scaling analysis based on the Monte Carlo simulations.
Our results are consistent with the Coulomb gas predictions for the critical O($n$) branch for $n < 2$ and the results
obtained by previous transfer matrix calculations.
For $n=2$, we find that the thermal exponent, the magnetic exponent and the fractal dimension of the largest
critical Eulerian bond component are different from those of the critical O(2) loop model. These results confirm that the cubic anisotropy
is marginal at $n=2$ but irrelevant for $n<2$.
%%%% 论文摘要
\end{minipage}
\end{center}

\begin{center}
\begin{minipage}{15.5cm}
\begin{minipage}[t]{2.3cm}{\bf Keywords:} \end{minipage}
\begin{minipage}[t]{13.1cm}
%%%%% 关键词
\end{minipage}\par\vglue8pt
{\bf PACS:} 05.50.+q, 64.60.Cn, 64.60.Fr, 75.10.Hk
%%% PACS 分类码
%% 查询网址：http://www.aip.org/pacs
\end{minipage}
\end{center}

\section{Introduction}
Generally, the Hamiltonian of the $n$-component spin cubic model\cite{henkcub1984,guocub2006} can be written as
\begin{equation}
{\mathcal H}/k_BT=-\sum\limits_{<i,j>}[K\vec{s}_i\cdot\vec{s}_j+M(\vec{s}_i\cdot\vec{s}_j)^2],
\end{equation}
where $T$ is the temperature, $k_B$ is the Boltzmann constant, and the sum on $<i,j>$ includes all pairs of nearest-neighbor (NN) sites.
The spin $\vec{s}_i$ is an $n$-component vector located on the $i$-th site, namely
$\vec{s}_i=(s_{i1},s_{i2},\cdots,s_{in})$, such that one and only one of the $n$ components has a nonzero value $\pm1$.
This model is also called `face-cubic model' because the spin can be regarded as a vector located at the
center of an $n$-dimensional hypercube, and pointing to the center of one of the faces of the hypercube.

This model combines the Potts degrees of freedom \cite{potts,wfypotts} with Ising degrees of freedom. Therefore the Hamiltonian can also be written as
\begin{equation}
{\mathcal H}/k_BT=-\sum\limits_{<i,j>}[Ks_is_j+M]\delta_{\sigma_i,\sigma_j},
\end{equation}
where $s_i=\pm1$ is an Ising spin, and $\sigma_i=1,2,\cdots,n$ is a Potts spin. The partition sum of the cubic model can be written as
\begin{eqnarray}
Z&=&\sum\limits_{\{s\},\{\sigma\}}e^{-{\mathcal H}/k_BT}=\sum\limits_{\{s\},\{\sigma\}}\prod\limits_{<i,j>}e^{(Ks_is_j+M)\delta_{\sigma_i,\sigma_j}} .
\end{eqnarray}
For the special case $\mbox{cosh}K=e^{-M}$,  this partition sum can be mapped to (see Ref. \cite{henkcub1984, guocub2006} for details)
\begin{equation}
Z=\sum\limits_{\{b\}}(nx)^{N_b}n^{N_c} \label{EBmodel},
\end{equation}
where $x=\frac{e^M \mbox{sinh}K}{n}$.
This model is defined in terms of bond variables that can take the values `absent' and `present'. The bond configuration ${\{b\}}$ is restricted to be Eulerian, which means that each
 site is connected to an {\it even} number of bonds.
 $N_b$ is the number of bonds, and $N_c$ is the number of components. Typically, a component is a group of sites connected by bonds,
but it can also be an isolated site.
The model may thus be called `Eulerian bond-cubic model', and one of its configurations is shown in Fig. \ref{dualsite}.
Furthermore, $n$ is no longer restricted to be an integer number in (\ref{EBmodel}), it can be any real number.

The Eulerian bond-cubic model has been studied by means of transfer matrix (TM)
calculations and a finite-size scaling analysis in Ref. \cite{guocub2006}.
In the region $1\le n\le 2$, critical points and three critical exponents were determined.
To further investigate the nature of the phase transition and the critical behavior, especially the geometric properties of critical configurations
of the model, one may also employ Monte Carlo simulations.
However, the problem arises to design an efficient Monte Carlo algorithm for this model, in view of the nonlocal
weight $n^{N_c}$.
A local update of the Metropolis type algorithm requires an nonlocal search to determine the
change of the number of components. This, together with the critical slowing-down, would make the simulation very
time-consuming in the critical region.

For this reason, we make use of the `coloring algorithm', which is proven to be useful for some models with nonlocal weights in their
partition sums, such as the random-cluster\cite{fk1,fk2} model and the O($n$) loop model\cite{onmodel}.
It was firstly proposed by Chayes and
Machta\cite{fkmc1,fkmc2}, and was originally combined with the Swendsen-Wang algorithm\cite{sw}  to simulate the
Potts model or the random-cluster model.
Ding {\it et al.} \cite{dingon2007} extended the application of the `coloring algorithm' to
the simulations of the O($n$) loop model on the honeycomb lattice using the Metropolis algorithm,
and greatly improved the efficiency of the algorithm.
Deng {\it et al.} \cite{dengon2007} further proposed two efficient cluster algorithms by combining the `coloring' trick with
the Swendsen-Wang algorithm for loop models: the algorithm 1 and the algorithm 2.
More applications of this `coloring' trick can be found in Ref. \cite{qianrc2005,dingon2009}.

By using the algorithm 2 \cite{dengon2007}, the Eulerian bond-cubic model (\ref{EBmodel}) has been preliminarily simulated.
However, only critical points were reported.
In this work, we develop a variant of the algorithm 1 to simulate the Eulerian bond-cubic model on the square lattice. We pay attention
not only to the thermodynamic properties of the model, e.g. the critical points, the thermal exponent and the magnetic exponent,
but also to geometric properties such as the fractal dimensions of the critical configurations.

The paper is arranged as follows: In Sec. \ref{algorithm}, we describe the cluster algorithm in
details for the Eulerian bond-cubic model. In Sec. \ref{vfss}, we describe the variables to be sampled and their finite-size scaling behaviors.
Our numerical results are given in Sec. \ref{results} and a summary is given in Sec. \ref{summary}.

\section{Algorithm}
\label{algorithm}
\begin{figure}
\begin{center}
\includegraphics[scale=0.4]{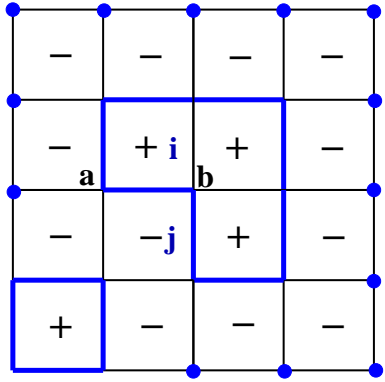}
\caption{A typical configuration of the Eulerian bond-cubic model and the definition of `dual sites', the sites $i$ and $j$ are the
`dual sites' of $a$ and $b$, reversely, $a$ and $b$ are also the `dual sites' of $i$ and $j$.}
\label{dualsite}
\end{center}
\end{figure}

The partition sum of the Eulerian bond-cubic model (\ref{EBmodel}) can be written as
\begin{equation}
Z=\sum\limits_{\{b\}}\sum\limits_{c_r=0}^{N_c} {N_c \choose c_r}(nx)^{N_b}(n-1)^{c_r}1^{c_g},
\end{equation}
where $c_r$ is the number of `red' components, and $c_g$ is the number of `green' components, with $c_r+c_g=n_c$.
Here, each Eulerian bond configuration is decomposed into a number of `colored-Eulerian-bond' configurations. After this
decomposition, the `coloring algorithm' can be applied in the procedure of the Monte Carlo
simulation of the model.
 We make use of the Ising spins sitting on the dual lattice to represent configurations: if two NN
 Ising spins $S_i$ and $S_j$ on the dual lattice are different, the corresponding edge on the original lattice
 between $S_i$ and $S_j$ is occupied by a bond. The Eulerian bonds are precisely the domain wall of the Ising spins,
 there is a two-to-one correspondence between the Ising-spin configurations $ {\{S\}}$ and the Eulerian bond configurations ${\{b\}}$,
 see Fig. \ref{dualsite} for example. In order to describe the algorithm more clearly, we define the
`dual sites': for an edge on the original lattice, the `dual sites' of the two sites connected by the edge are the two sites on the
 dual lattice that sit at the two sides of the edge, reversely, the two sites connected by an edge on the original lattice are also
 the `dual sites' of the two sites that sit at the two sides of the edge. See Fig. \ref{dualsite} for example.
Then the Swendsen-Wang type algorithm with `coloring' trick for the Eulerian bond-cubic model can be described as:
\begin{enumerate}
\item Start from an arbitrary Ising-spin configuration, which corresponds to an Eulerian bond configuration
      on the original lattice.
\item Set color to the components of the Eulerian bond configuration.
      For each component on the original lattice, set it as
      green (active) with probability $p=1/n$, or red (inactive) with probability $1-p$.
\item Construct the Swendsen-Wang clusters by placing percolation bonds on the dual lattice.

%     For each pair of the NN sites $i$ and $j$ on the dual lattice,

      \begin{enumerate}
      \item[$\bullet$] For each pair of NN sites $i$ and $j$ on the dual lattice, a percolation bond is placed between them with probability $p=1$
                       if not all of the colors of their dual sites are green.
      \item[$\bullet$] If all of the colors of their dual sites are green and $S_i=S_j$, connect them by a percolation bond with probability $p=1-nx$;
                       otherwise, let the edge be vacant.
      \item[$\bullet$] Each pair of NN sites on the dual lattice is considered to be in the same cluster if there is a percolation bond between them.
                       These percolation clusters are called Swendsen-Wang clusters.
      \end{enumerate}
\item Flip every Swendsen-Wang cluster with probability $p=1/2$.
\item Sample the variables of interest, erase the colors and restart at step 2.
\end{enumerate}
The algorithm can be modified to be Wolff-type\cite{wolff}  by constructing only one cluster which is then flipped with probability $p=1$ in step 3 and 4.

%In the step 3 and 4, if one constructs only one cluster and flip it with probability $p=1$, the algorithm becomes a Wolff-type
% algorithm\cite{wolff}.

\section{Sampled variables and their finite-size scaling behavior}
\label{vfss}
\begin{figure}[htbp]
\begin{center}
{\includegraphics[scale=0.4]{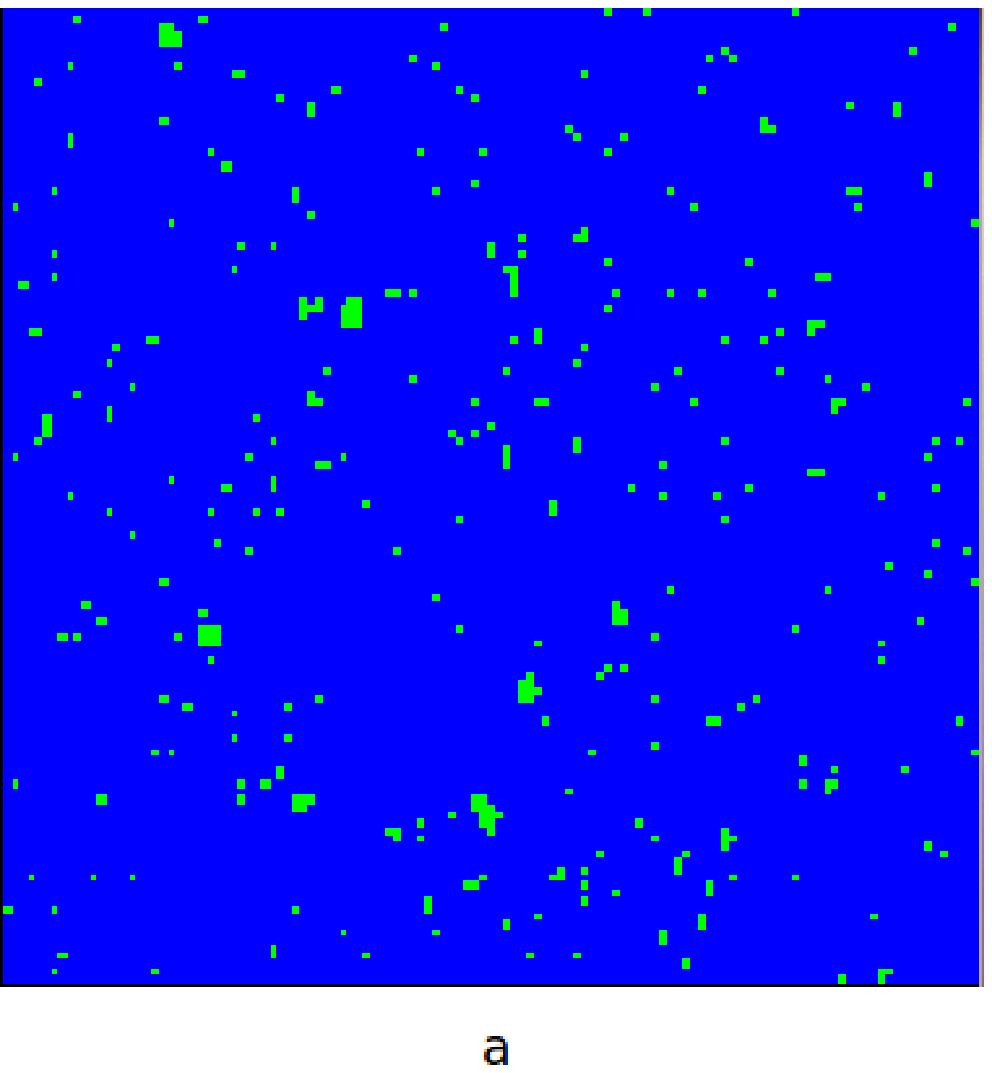}}
{\includegraphics[scale=0.4]{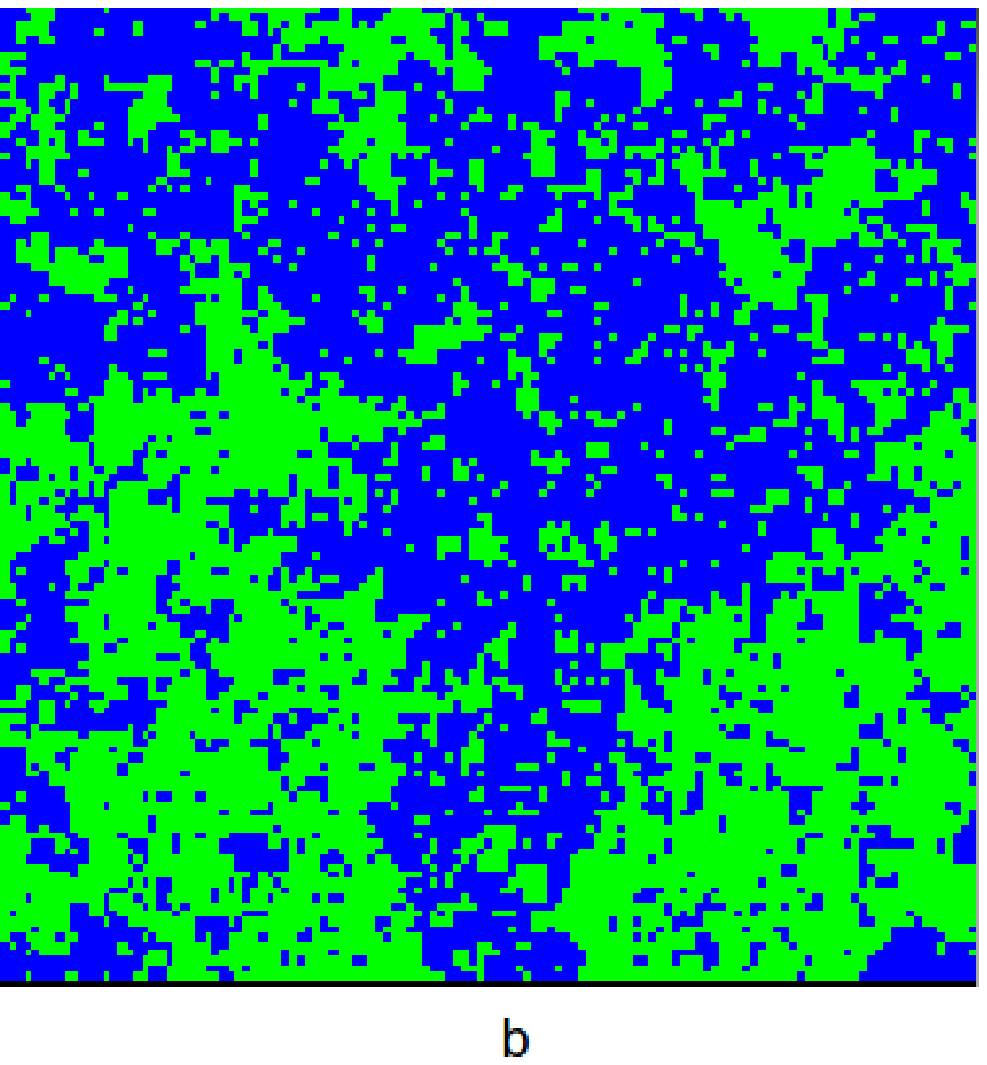}}
{\includegraphics[scale=0.4]{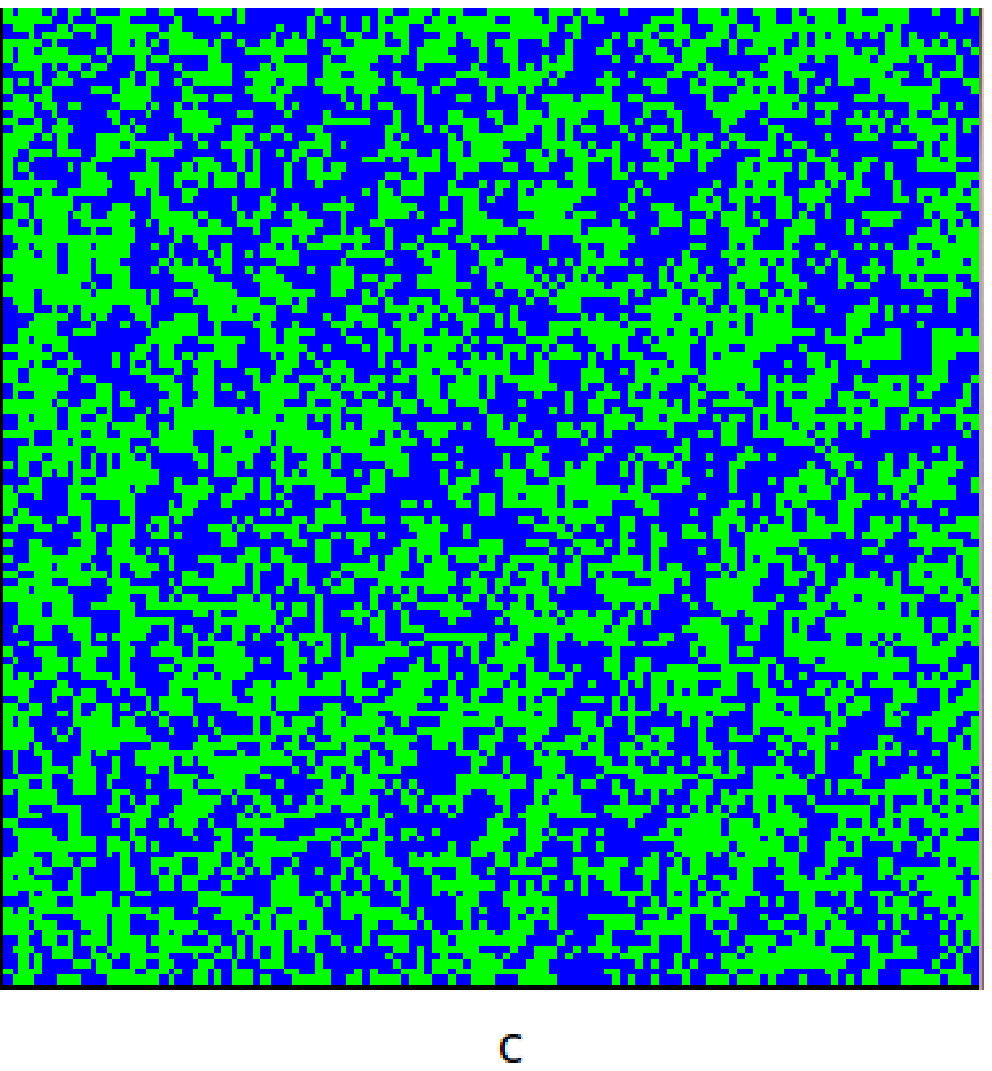}}
\caption{(Color online) Three typical configurations for $n=1.5$ Eulerian bond-cubic model with system size $L=128$: (a) at high
temperature $x=0.3$, (b) at critical point $x=0.444245$, (c) at low temperature $x=0.6$. Green means the spin on the dual lattice
 is `+', blue means `-'; the domain wall of the Ising clusters are exactly the Eulerian bond configurations.
}
\label{cfg}
\end{center}
\end{figure}

%For the Eulerian bond-cubic model with given $n$, when $x$ is small, a typical configuration has only a few small components, see Fig. \ref{cfg}(a);
%when $x$ becomes larger, the typical configuration has more bonds and bigger components; when $x$ reaches or exceeds the
% critical point $x_c$, the typical configuration will have a component that spans the whole lattice, see Fig. \ref{cfg}(b) and (c).
A typical high-temperature (small $x$) configuration of the Eulerian bond-cubic model with a given $n$ has
only a few small components, as shown in Fig. \ref{cfg}(a).
When $x$ becomes larger, the typical configuration has more bonds and bigger components. A component that spans the whole lattice will emerge
when $x$ reaches or exceeds the critical point $x_c$, see Fig. \ref{cfg}(b) and (c).

\begin{figure}[htbp]
\begin{center}
\includegraphics[scale=1.0]{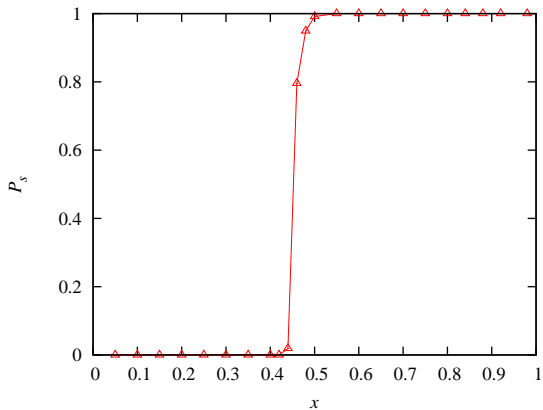}
\caption{The percolation probability $P_s$ versus $x$ for $n=1.5$ Eulerian bond-cubic model with system size $L=256$.}
\label{pst15}
\end{center}
\end{figure}

The behavior of the components in these configurations is very similar to that of clusters in percolation phenomena\cite{stauffer}, so we call the spanning
component a `percolating component'.
% We can describe the phase transition of the model in the language of percolation;
The percolation probability $P_s$ is defined as
\begin{eqnarray}
P_s=\langle P \rangle \ ,
\end{eqnarray}
where $P$ is 1 if there exists  a percolating component on the configuration, 0 otherwise. $\langle \cdots \rangle$ means the average over the canonical ensemble.
For an infinite system, $P_s$ is 1 for $x>x_c$ and $0$ for $x<x_c$, which is a $\Theta$ function. However, for a finite system, the value of
$P_s$ changes continuously when $x$ passes $x_c$, as shown in Fig. \ref{pst15}.

\begin{figure}[htbp]
\begin{center}
\includegraphics[scale=1.0]{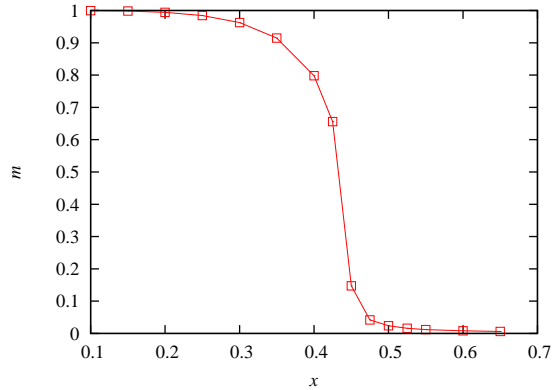}
\caption{The magnetization $m$ versus $x$ for $n=1.5$ Eulerian bond-cubic model with system size $L=256$.}
\label{m15}
\end{center}
\end{figure}

On the other hand, the phase transition of this model can be described by in terms of the Ising spins on the dual lattice. When $x$ is small,
most Ising spins have the same sign, thus the system is in a long-range ordered state (ferromagnetic) and has a nonzero
spontaneous magnetization, as shown in Fig. \ref{cfg}(a). When $x$ is large, the Ising spins will be in a disordered state (paramagnetic),
and the magnetization will be zero, as shown in Fig. \ref{cfg}(c). The phase transition of the system is a ferromagnetic one. Concretely,
the magnetization $m$ is defined as
\begin{eqnarray}
m&=&\langle{\mathcal M} \rangle,
\end{eqnarray}
with
\begin{eqnarray}
{\mathcal M}&=& \bigg|\frac{\sum\limits_iS_i}{V}\bigg|,
\end{eqnarray}
where $V=L^d$ is the system volume and $d$ is the dimension of the lattice. In the current paper, $d=2$. Figure \ref{m15} shows the magnetization $m$ versus
 $x$ for $n=1.5$ Eulerian bond-cubic model with system size $L=256$. Figure \ref{cfg}, \ref{pst15} and \ref{m15}
give a general description of the phase transition of the model.

The critical point can be determined by the percolation probability $P_s$ and the Binder ratio of $m$
\begin{eqnarray}
Q&=&\frac{\langle {\mathcal M}^2\rangle^2}{\langle{\mathcal M}^4\rangle} .
\end{eqnarray}
 According to the renormalization theory, the percolation probability $P_s$, the magnetization $m$ and the Binder ratio $Q$
display the following finite-size scaling behavior\cite{fssa,fssb}
\begin{eqnarray}
P_s&=&P_s^{(0)}+a_1(x-x_c)L^{y_t}+a_2(x-x_c)^2L^{2y_t}+\cdots+b_1L^{y_1}+b_2L^{y_2}+\cdots, \label{psfss}\\
m&=&L^{y_m-d}[m_0+a_1(x-x_c)L^{y_t}+a_2(x-x_c)^2L^{2y_t}+\dots+b_1L^{y_1}+b_2L^{y_2}+\cdots\label{mfss}], \\
Q&=&Q_0+a_1(x-x_c)L^{y_t}+a_2(x-x_c)^2L^{2y_t}+\cdots+b_1L^{y_1}+b_2L^{y_2}+\cdots, \label{Qfss}
\end{eqnarray}
where $y_t>0$ and $y_m>0$ are the thermal exponent and magnetic exponent respectively, and $y_1$, $y_2, \cdots, <0$ are the correction-to-scaling
 exponents. $x_c$ is the critical point and $m_0$, $a_i$, $b_i$ are unknown constants.
We can see that $P_s$ and $Q$ have a similar finite-size scaling behavior.  Both (\ref{psfss}) and (\ref{Qfss})
can be used to determine the critical point and the thermal exponent, as will be shown in more detail in Sec. \ref{results}.

At the critical point $x_c$, (\ref{mfss}) reduces to
\begin{equation}
m=L^{y_m-d}(m_0+b_1L^{y_1}+b_2L^{y_2}+\cdots) \label{mfssc}.
\end{equation}
 Both (\ref{mfss}) and (\ref{mfssc}) can be used to determine the magnetic exponent $y_m$, however, (\ref{mfssc})
 is used more in practice because it has fewer parameters thus is more convenient in a data analysis.

Besides the critical exponents $y_t$ and $y_m$, we are also interested in the fractal structure of the critical configurations.
There are two fractals on the critical configuration: the largest Eulerian bond components and the largest `Ising cluster'.
The Ising cluster is defined as a group of NN Ising spins in the same sign.
%For critical Eulerian-bond configurations, the largest critical component is a fractal. The Ising-spin configurations on the dual
%lattice are in a two-to-one correspondence to the Eulerian-bond configurations, and we define the `Ising cluster': two nearest-neighboring
%Ising spins belong to the same Ising cluster, if they have the same sign.
%At the critical point, the largest Ising cluster is also a fractal.
We define the percolation strength $P^{\alpha}_{\infty}$ and the average size $\chi^{\alpha}$ based on the components and Ising clusters.
\begin{eqnarray}
P^{\alpha}_{\infty}&=&\bigg< \frac{c_{\infty}}{\sum\limits_i c_i}\bigg> ,\\
\chi^{\alpha}&=&\bigg<\frac{\sum\limits_i c_i^2}{\sum\limits_i c_i}\bigg>,
\end{eqnarray}
where the superscript $\alpha=s$, $b$ or $c$. For $\alpha=s$, $c_i$ is the number of sites in the $i$-th component;
for $\alpha=b$, $c_i$ is the number of bonds in the $i$-th component; and for $\alpha=c$, $c_i$ is the number of Ising spins
in the $i$-th Ising cluster. $c_{\infty}$ is the size of the largest component or Ising cluster on the configuration.
The subscript $\infty$ is used because only the largest component or the largest Ising cluster may have
 an nonzero fraction comparing to the system volume in the thermodynamic limit. The $P^{\alpha}_{\infty}$ can be
 considered as the order parameter of the phase transition, playing the role of the magnetization in the Ising model. The Greek
 letter $\chi$ is used to denote the average size,
 because it has the property that is very similar to the magnetic susceptibility of the Ising model.

$P^{\alpha}_{\infty}$ and $\chi^{\alpha}$ have the finite-size scaling behaviors similar to that of the magnetization:
\begin{eqnarray}
P^{\alpha}_{\infty}&=&L^{y-d}( P^{(0)}_{\infty}+a_1(x-x_c)L^{y_t}+a_2(x-x_c)^2L^{2y_t}+\cdots+b_1L^{y_1}+b_2L^{y_2}+\cdots),\label{pfss} \\
\chi^{\alpha}&=&L^{2y-d}( \chi_0+a_1(x-x_c)L^{y_t}+a_2(x-x_c)^2L^{2y_t}+\cdots+b_1L^{y_1}+b_2L^{y_2}+\cdots).\label{xfss}
\end{eqnarray}
When $\alpha=s$ or $b$, $y=y_H$ is the fractal dimension of the largest component of the critical
Eulerian bond configuration, which can be viewed as the hull of the largest Ising cluster;
when $\alpha=c$, then $y=y_c$ is the fractal dimension of the largest critical Ising cluster.
$P^{(0)}_{\infty}$, $\chi^0$, $a_i$ and $b_i$ are unknown constants, and the $y_i<0$ are irrelevant exponents. At the critical
point $x_c$, (\ref{pfss}) and (\ref{xfss}) reduce to
\begin{eqnarray}
P^{\alpha}_{\infty}&=&L^{y-d}(P^{(0)}_{\infty}+b_1L^{y_1}+b_2L^{y_2}+\cdots), \label{pfssc}\\
\chi^{\alpha}&=&L^{2y-d}(\chi_0+b_1L^{y_1}+b_2L^{y_2}+\cdots). \label{xfssc}
\end{eqnarray}
 We will use (\ref{pfssc}) and (\ref{xfssc}) to determine $y_H$  and $y_c$.
\section{Results}
\label{results}
\begin{figure}[htbp]
\begin{center}
\includegraphics[scale=1.0]{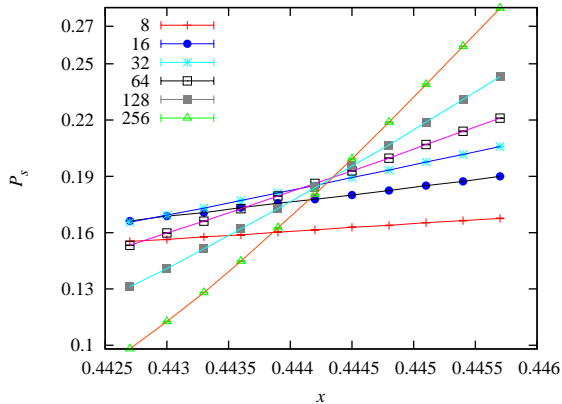}
\caption{The percolation probability $P_s$ versus $x$ for various system sizes of the $n=1.5$ Eulerian bond-cubic model.}
\label{ps1.5}
\end{center}
\end{figure}

Using the  Swendsen-Wang type or Wolff type algorithm described in Sec. \ref{algorithm}, we do Monte Carlo simulations
 for the Eulerian bond-cubic model on the square lattice. We apply $10^4$ Swendsen-Wang/Wolff cycles to equilibrate the system,
 and average over $2\times 10^7$ samples, where each sample is taken after every 3 cycles. The sizes of the simulated systems range
 from $L$=8 to $L$=256. Figure \ref{ps1.5} shows part of the data of $P_s$ versus $x$ for $n=1.5$ Eulerian bond-cubic model near the critical
 point $x_c$. We fit the data according to (\ref{psfss}) using the nonlinear Levenberg-Marqurdt least-squares algorithm, which yields the thermal exponent
$y_t=0.747(6)$ and the critical point $x_c=0.4443(2)$.

\begin{figure}[htpb]
\begin{center}
\includegraphics[scale=1.0]{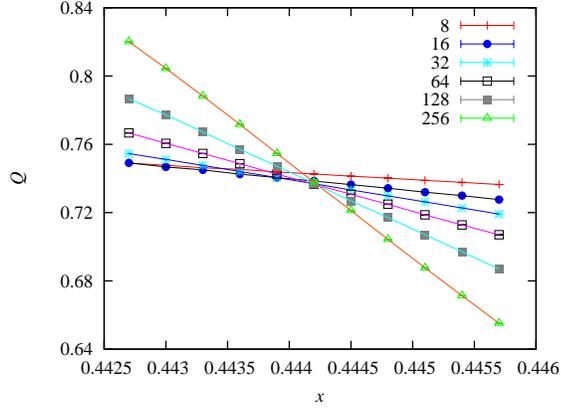}
\caption{The Binder ratio $Q$ versus  $x$ for various system sizes of the $n=1.5$ Eulerian bond-cubic model.}
\label{q1.5}
\end{center}
\end{figure}

Figure \ref{q1.5} shows part of the data of $Q$ versus $x$ for $n=1.5$ Eulerian bond-cubic model. Fitting the data according to (\ref{Qfss}), we obtain the
thermal exponent $y_t=0.748(3)$ and the critical point  $x_c=0.444245(8)$. We can see that results for $y_t$ and $x_c$ from the fit to the data
 for $P_s$ agree well with the ones from the fit to the data for $Q$.

According to Coulomb gas theory, for $1 \le n <2$, the Eulerian bond-cubic model belongs to the same universality class of the critical
O($n$) loop model, with critical exponents \cite{nienhuis1982,nienhuis1984,SD,DS}
\begin{eqnarray}
y_t&=&4-\frac{4}{g} \label{yt},\\
y_m&=&3-\frac{3}{2g} \label{ym},\\
y_H&=&1+\frac{1}{2g} \label{yH},\\
y_c&=&1+\frac{g}{2}+\frac{3}{8g},\label{yc}
\end{eqnarray}
with $n=-2\mbox{cos}(\pi g)$, $1\le g\le2$. $g$ is the Coulomb gas coupling constant.
According to (\ref{yt}), we have $y_t=0.748109\cdots$ for $n=1.5$ Eulerian bond-cubic model, which is consistent with the
numerical result.
For the critical point of the Eulerian bond-cubic model on the square lattice, there is no exact result except
for the cases $n=1$ and $n=2$. $x_c$ is $\sqrt{2}-1$ for the case $n=1$ and 0.5 for the case $n=2$\cite{guocub2006}.
The critical points and some critical exponents of the Eulerian bond-cubic model
are also determined by a finite-size scaling analysis based on numerical TM calculations
in Ref. \cite{guocub2006}, where it is found
$x_c=0.44424(1)$ and $y_t=0.749(2)$ for the $n=1.5$ Eulerian bond-cubic model,
which are in good agreement with our Monte Carlo results. % are also consistent with those of TM's.

We have also simulated the cases $n=1.0$, $1.1$, $1.25$, $1.7$, $1.75$, and $2.0$, both the Coulomb gas predictions for the critical O($n$) branch
and the numerical results for $y_t$ and $x_c$ are summarized in Tab. \ref{ytxc}.
\begin{table}[htbp]
\setlength{\tabcolsep}{2pt}
\begin{center}
\caption{Thermal exponents and critical points of the Eulerian bond-cubic model on the square lattice. T = theoretical predictions of the critical
O($n$) branch,
$P_s$ = Monte Carlo results from the fit to the data for $P_s$, $Q$ = Monte Carlo results from the fit to the data for $Q$, TM = results
calculated by TM\cite{guocub2006}.}
\begin{tabular}{|l|l|l|l|l|l|l|l|l|l|}
\hline
 $n$&$y_t$(T) &$y_t$($P_s$)&$y_t$($Q$)& $y_t$(TM) &$x_c$($P_s$)&$x_c$($Q$)&$x_c$(TM) &$Q_0$ &$P_s^{(0)}$\\
\hline
 1.0 &1.000000&1.01(2) &0.998(2)&1.000000(1)&0.41422(1)&0.414214(2)&0.4142135(1)&0.8560(2)&0.0651(3)\\
 1.1 &0.957313&0.959(3)&0.957(3)&0.9572(5)  &0.41916(1)&0.419155(4)&0.419154(2) &0.8359(2)&0.0812(4)\\
 1.25&0.887399&0.890(4)&0.888(2)&-          &0.42741(1)&0.427404(3)&-           &0.8022(1)&0.110(1)\\
 1.5 &0.748109&0.747(6)&0.748(3)&0.749(2)   &0.4443(2) &0.444245(8)&0.44424(1)  &0.7454(3)&0.1904(5)\\
 1.7 &0.600379&0.598(4)&0.604(6)&0.595(5)   &0.4624(3) &0.46213(2) &0.46214(1)  &0.6686(3)&0.294(2)\\
 1.75&0.552482&0.554(2)&0.57(3) &-          &0.4679(4) &0.46754(2) &-           &0.649(2) &0.329(1)\\
 2.0 &0.000000&0.52(3) &0.48(3) &0.50000(1) &0.4998(3) &0.50001(1) &0.5000000(1)&0.550(2) &0.554(5)\\
% 2.0 &0.500000&0.52(3) &0.48(3) &0.50000(1) &0.4998(3) &0.50001(1) &0.5000000(1)&0.550(2) &0.554(5)\\
\hline
\end{tabular}
\label{ytxc}
\end{center}
\end{table}

One should pay more attention to the case of $n=2$. The thermal exponent $y_t$ is 0 for the critical branch of O(2) loop
model, while it is 0.5 for $n=2$ Eulerian bond-cubic model. For the case $n=2$, the Eulerian bond-cubic model reduces
to a special case of the Ashkin-Teller model\cite{nienhuis1984,denNijs, Knops, Baxter}. Our estimations of $y_t$ agree with the exact results
$y_t=0.5$ for the Ashkin-Teller model.
%For a detailed discussion of this problem, see Ref. \onlinecite{guocub2006}.

At the critical point, we sampled the magnetization $m$, percolation strength $P^c_{\infty}$, $P^s_{\infty}$, $P^b_{\infty}$
and the average cluster sizes $\chi^c$, $\chi^s$, $\chi^b$.

\begin{figure}
\begin{center}
\includegraphics[scale=1.0]{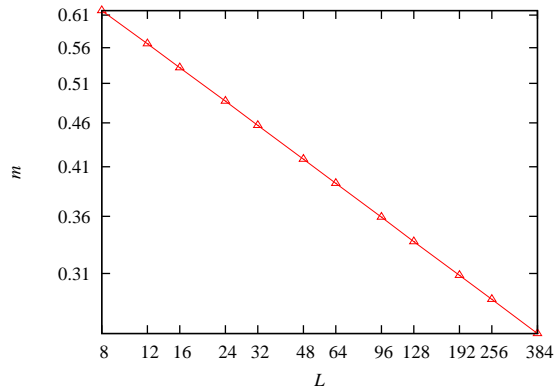}
\caption{The log-log plot of the magnetization $m$ versus system size $L$ for $n=1.5$ Eulerian bond-cubic model. The error bars are
much smaller than the size of the data points.}
\label{m1.5}
\end{center}
\end{figure}

The log-log plot of the data for magnetization $m$ versus system size $L$ for $n=1.5$  Eulerian bond-cubic model at the critical point
is shown in Fig. \ref{m1.5}. Fitting the data according to (\ref{mfssc}), we obtain the magnetic exponent $y_m=1.7803(3)$, which is
consistent with the theoretical prediction $y_m=1.78054\cdots$, given by (\ref{ym}).

\begin{figure}
\begin{center}
\includegraphics[scale=1.0]{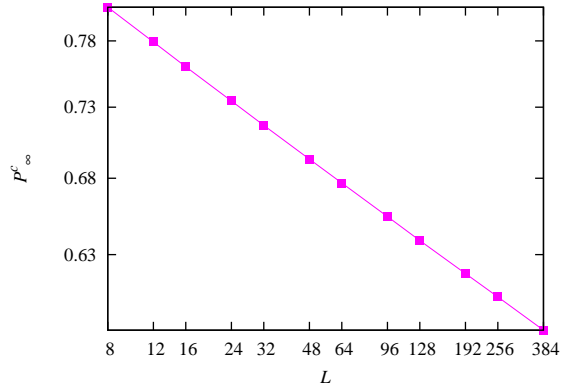}
\caption{The log-log plot of the percolation strength $P^c_{\infty}$ versus system size $L$ for $n=1.5$ Eulerian bond-cubic model. The error bars are much smaller than the size of the data points.}
\label{p1.5}
\end{center}
\end{figure}

\begin{figure}
\begin{center}
\includegraphics[scale=1.0]{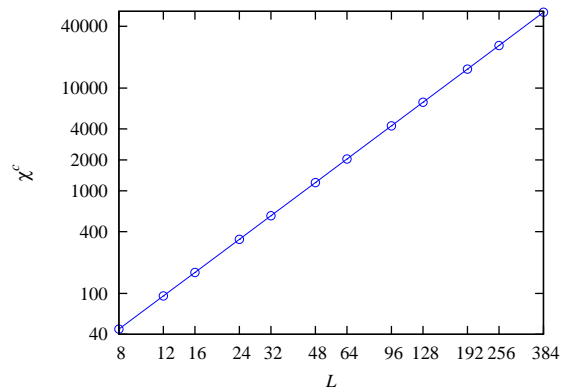}
\caption{The log-log plot of the average size $\chi^c$ versus system size $L$ for $n=1.5$ Eulerian bond-cubic model. The error bars are much smaller than the size of the data points.}
\label{xi1.5}
\end{center}
\end{figure}

The log-log plot of the data for $P^c_{\infty}$ and $\chi^c$ versus $L$ are shown
in Fig. \ref{p1.5} and Fig. \ref{xi1.5} respectively. Fitting the data, we obtain the fractal dimension of the largest critical Ising
 cluster $y_c=1.9195(10)$ (fit from $P^c_{\infty}$) and $y_c=1.9195(11)$ (fit from $\chi^c$), which
are consistent with the theoretical prediction $y_c=1.91989\cdots$, given by (\ref{yc}).

 Fitting the data of $P^s_{\infty}$, $\chi^s$, $P^b_{\infty}$ and $\chi^b$,
 we obtain the fractal dimension of the largest critical component $y_H=1.410(5)$ (fit from $P^s_{\infty}$), $y_H=1.405(4)$
(fit from $\chi^s$), $y_H=1.410(4)$ (fit from $P^b_{\infty}$), and $y_H=1.41(1)$ (fit from $\chi^b$). All of them are
consistent with the theoretical prediction $y_H=1.4064\cdots$, given by (\ref{yH}).

All the numerical results and the theoretical predictions of $y_m$, $y_c$ and $y_H$ are listed in Tab. \ref{ymyc} and Tab. \ref{YH}.
From these tables, we can see that the fractal dimension $y_c$ of the largest critical Ising cluster of the $n=2$ Eulerian bond-cubic model is
the same as the critical O(2) value. However, the fractal dimension of the largest critical Eulerian bond
component is $y_H=1.625$, which is obviously different from the critical O(2) value
$y_H=1.5$, given by (\ref{yH}). Also the magnetic exponent $y_m$ is different from the critical O(2) value 1.5.
These results agree well with the exact results $y_m=1.625=y_H$ for the Ashkin-Teller model \cite{denNijs, Knops, Baxter}, and again
show the difference between the cubic symmetry and the O($n$) symmetry in the case $n=2$,
when the cubic anisotropy becomes marginally relevant.

\begin{table}[htbp]
\begin{center}
\caption{The magnetic exponents $y_m$ and fractal dimension $y_c$ of the Eulerian bond-cubic model on the square lattice.
T = theoretical predictions of the critical O($n$) branch, MC = Monte Carlo results, $P^c_{\infty}$ = Monte Carlo results from the fit to the data for $P^c_{\infty}$,
$\chi^c$ = Monte Carlo results from the fit to the data for $\chi^c$, TM = results calculated by TM\cite{guocub2006}. }
\begin{tabular}{|l|l|l|l|l|l|l|l|l|l|}
\hline
 $n$&$y_m$(T) &$y_m$(MC) & $y_m$(TM) &$y_c$(T) & $y_c$($P^c_{\infty}$) &$y_c$($\chi^c$)\\
\hline
 1.0 &1.875   &1.8751(2) &1.87501(1) &1.94792 &1.9476(3)&1.9477(3)  \\
 1.1 &1.85899 &1.8590(1) &1.85895(5) &1.94257 &1.9422(3)&1.9423(3)  \\
 1.25&1.83277 &1.832(1)  &-          &1.93436 &1.933(2) &1.933(2)   \\
 1.5 &1.78054 &1.7803(3) &1.7805(5)  &1.91989 &1.9195(10)&1.9195(11)\\
 1.7 &1.72514 &1.724(1)  &1.726(2)   &1.90702 &1.906(1) &1.909(3)   \\
 1.75&1.70786 &1.709(2)  &-          &1.90347 &1.902(2) &1.90(1)    \\
 2.0 &1.5     &1.6249(4) &1.62500(1) &1.875   &1.875(1) &1.8750(1)  \\
% 2.0 &1.625   &1.6249(4) &1.62500(1) &1.875   &1.875(1) &1.8750(1)  \\
\hline
\end{tabular}
\label{ymyc}
\end{center}
\end{table}

\begin{table}[htbp]
\begin{center}
\caption{ The fractal dimension $y_H$ of the Eulerian bond-cubic model on the square lattice.
T = theoretical predictions of critical O($n$) branch, $P^s_{\infty}$ = Monte Carlo results from the fit to the data for $P^s_{\infty}$,
$\chi^s$ = Monte Carlo results from the fit to the data for $\chi^s$, and so forth.}
\begin{tabular}{|l|l|l|l|l|l|}
\hline
 $n$ & $y_H$(T)& $y_H$($P^s_{\infty}$)&$y_H$($\chi^s$)&$y_H$($P^b_{\infty}$) &$y_H$($\chi^b$)  \\
\hline
 1.0 &1.375 &1.376(1) &1.373(4) &1.378(4)  &1.372(4)\\
 1.1 &1.3803&1.3806(9) &1.380(5)&1.382(4)  &1.379(3)\\
 1.25&1.3891&1.391(2) &1.390(2) &1.393(4)  &1.389(1)\\
 1.5 &1.4064&1.410(5) &1.405(4) &1.410(4)  &1.41(1)\\
 1.7 &1.4249&1.426(4) &1.43(2)  &1.42(1)   &1.430(6)\\
 1.75&1.4307&1.43(1)  &1.431(12)&1.427(11) &1.43(1)\\
 2.0 &1.5   &1.625(1) &1.6253(5)&1.625(1)  &1.624(1)\\
\hline
\end{tabular}
\label{YH}
\end{center}
\end{table}
% y_c

\section{Summary}
\label{summary}
We simulated the Eulerian bond-cubic model on the square lattice using an efficient cluster algorithm.
Two fractal dimensions of the critical configurations as well as the critical points, the thermal and magnetic exponents
of the model are determined by means of a finite-size scaling analysis. The two fractal dimensions
are for the first time obtained numerically.
The estimations of the critical points and the thermal and magnetic exponents are in good agreement with those obtained by means of TM
calculations \cite{guocub2006} for several values of $n$, including the case $n=2$.
Our results for all critical exponents are consistent with the Coulomb gas predictions of the critical O($n$) branch for $n<2$.
But, for $n=2$, the thermal exponent, the magnetic exponent and the fractal dimensions $y_H$ are different from the critical O(2) values,
and the model reduces to a special case of the Ashkin-Teller model.
Our study confirms that the phase transition of the Eulerian bond-cubic model belongs to the critical O($n$) universality class for
$n<2$. The cubic anisotropy is irrelevant for $n<2$, but becomes marginal when $n=2$.
%But for the $n=2$ Eulerian bond-cubic model, we find its thermal exponent and fractal dimension
%of the largest critical Eulerian bond component are different from those of the O($2$) loop model,
%which reveals that

\section*{Acknowledgment}
This research is supported by the High Performance Scientific
Computing Center(HSCC) of the Beijing Normal University.
%\appendix*

%\end{CJK*}
\end{document}